\documentclass[pre,aps,twocolumn,floatfix]{revtex4}
\usepackage[dvips]{color,graphicx}
\usepackage{amsmath}
\usepackage{amsfonts}
\usepackage{bm}
\begin{document}

\date{\today}
\title{  Monte Carlo study of degenerate groundstates and residual entropy \\ in a frustrated honeycomb lattice Ising model }
\author{Shawn Andrews}
\affiliation{Department of Applied Mathematics, University of Waterloo, Ontario, N2L 3G1, Canada} 
\author{Stephen Inglis}
\affiliation{Department of Physics and Astronomy, University of Waterloo, Ontario, N2L 3G1, Canada} 
\author{Roger G. Melko}
\affiliation{Department of Physics and Astronomy, University of Waterloo, Ontario, N2L 3G1, Canada} 
\author{Hans  De Sterck}
\affiliation{Department of Applied Mathematics, University of Waterloo, Ontario, N2L 3G1, Canada}

\begin{abstract}
We study a classical fully-frustrated honeycomb lattice Ising model using Markov chain Monte Carlo methods and exact calculations .  The Hamiltonian realizes a degenerate ground state manifold of equal-energy states, where each hexagonal plaquette of the lattice has one and only one unsatisfied bond, with an extensive residual entropy that grows as the number of spins $N$.  Traditional single-spin flip Monte Carlo methods fail to sample all possible spin configurations in this ground state efficiently, due to their separation by large energy barriers.  We develop a non-local ``chain-flip'' algorithm that solves this problem, and demonstrate its effectiveness on the Ising Hamiltonian with and without perturbative interactions.  The two perturbations considered are a slightly weakened bond, and an external magnetic field $h$.   For some cases, the chain-flip move is necessary for the simulation to find an ordered ground state.   In the case of the magnetic field, two magnetized ground states with non-extensive entropy are found, and two special values of $h$ exist where the residual entropy again becomes extensive, scaling proportional to $N\ln \phi$, where $\phi$ is the golden ratio.

\end{abstract}
\maketitle


\section{Introduction}

Ising frustration is a common ingredient in spin models designed to search for and study exotic physics.    The prototypical example is the well-known triangular lattice antiferromagnetic (AFM) Ising model,
\begin{equation}
H = J \sum_{\langle ij \rangle} S^z_i S^z_j, \label{Ising}
\end{equation}
which admits a ground state without long-range order (with power-law spin correlations), where an extensive number of degenerate (equal-energy) configurations causes a residual ($T=0$) entropy \cite{Wannier,Hot}.  This classical ``manifold'' of ground states is the fertile foundation from which one expects novel or exotic order to spring.  For example, the triangular lattice AFM Ising Hamiltonian, perturbed with a quantum transverse field, undergoes order-by-disorder to realize a long-range ordered quantum dimer state \cite{Moessner1}.   If instead the perturbation is a nearest-neighbor in-plane ferromagnetic quantum exchange, the system reveals an exotic ``supersolid'' phase with coexisting diagonal and off-diagonal long-range order \cite{Wessel_SS, Kedar_SS,Melko_SS}.  Ultimately, one would like to construct a spin Hamiltonian that is a true $T=0$ quantum paramagnet,  spin-liquid \cite{SL_Anderson}, or resonating valence-bond phase \cite{RVB, QDM_Mis}.  This could open a new window on our understanding of the world of deconfinement, quantum number fractionalization, and topological order \cite{Wen}, a role in which frustrated interactions are sure to play.

In the case of purely classical systems, perturbations to frustrated Ising Hamiltonian are of utmost importance to actual material physics, as is well documented in the spin ices \cite{SpinIce} -- rare-earth titanates that realize to a very close approximation Ising models on the frustrated pyrochlore lattice.  In addition to the Ising exchange, in these materials the dipolar interaction strength is significantly large, and has been shown to be a critical ingredient in the realization of the spin ice state \cite{Byron,Isakov1} as well as the prediction for long-range order \cite{LRO}.

\begin{figure}
 \includegraphics[width=3in]{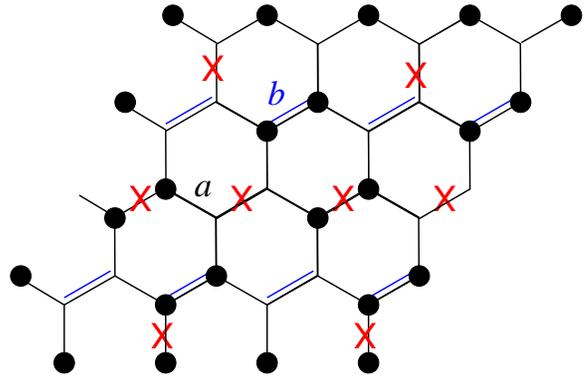}
 \caption{\label{latt_FFH}
(Color online) A fully-frustrated honeycomb lattice.  Single lines (labelled $a$) represent antiferromagnetic bond interactions, and double (blue) lines (labelled $b$) represent ferromagnetic interactions. The lattice illustrated has $N=2\times L \times L$ sites, with $L=4$, and periodic boundary conditions.  Dots illustrate one possible ground-state configuration of the Ising model, with black dots representing $S^z$ spin-up, and empty sites spin-down.   A (red) ex denotes each unsatisfied bond.}  
\end{figure}

A special class of classical Ising model is of particular theoretical interest due to the ability to map their ground states to hard-core dimer models, in which dimers live on the {\it bonds} of the respective dual lattice \cite{Moessner1,F_dimer}.   The simplest frustrated dimer model is the classical triangular lattice dimer model \cite{Fendley};  extensions of this prototypical example are known to harbor interesting physical phenomena.   The goundstate of this model is one where each {\it site} of the triangular lattice has one and only one dimer emanating from it.
Although no long-range order exists in this ground state, when constrained to a torus the model admits configurations which may be categorized into four distinct topological sectors \cite{F_dimer}.  Quantum extensions of this model promote, among other things, a short-ranged resonating-valence bond phase with deconfined fractional excitations (spinons) \cite{RVB}.  The quantum dimer model on the triangular lattice has also been used to motivate the design of topologically protected qbits \cite{Qbit}.  Clearly, such models are ideal playgrounds for the study of the exotic physics mentioned above.

The ground state of the classical dimer model on the triangular lattice maps to the so-called {\it fully-frustrated} (FF) honeycomb lattice Ising model \cite{Moessner1,F_dimer}.  In that model, antiferromagnetically interacting spins are placed on the sites of a honeycomb lattice (labelled $a$ in Fig.~\ref{latt_FFH}), with the exception that {\it one} bond per hexagon has an exchange of opposite sign, i.e.~a ferromagnetically interacting nearest-neighbor pair (labelled $b$):
\begin{equation}
H =  J\sum_{\langle ij \rangle} (-1)^{\delta_{\langle i j \rangle, b}} S^z_i S^z_j.  \label{FFh}
\end{equation}
Here, $\delta_{\langle i j \rangle, b}$ is a delta function, with value 0 on bonds $a$ and unity on bonds $b$.
In this paper, we use the convention $S^z_i = \pm 1/2$.  The ground state of this model is one where each hexagon of the honeycomb lattice has one and only one unsatisfied bond.  If this unsatisfied bond is mapped to represent a dimer on the dual (triangular) lattice, one immediately sees that the 
spin configurations in the ground state of the FF honeycomb lattice Ising model can be matched to the ground state of the classical  dimer model on the triangular lattice (a close-packed model of dimers with hard cores), which has one and only one dimer connected to each site \cite{Fendley}. 
This ground state does not have a long-range order; rather, it is an extensive manifold of equal-energy disordered states, which produce a residual entropy of $S\approx 0.214$ per spin \cite{Fendley,Wolff}.  

We note that the equivalency of the ground-states of the two models (FF honeycomb Ising and triangular dimer) is true for many choices of the pattern of the FM bonds (not just the one in  Fig.~\ref{latt_FFH}).  In other words, there is a ``gauge freedom'' for which bonds we call ferromagnetic and which we call antiferromagnetic, as occurs with similar mixed-bond models \cite{Moessner1}.  Other patterns for the FM bonds could be chosen in Fig.~\ref{latt_FFH} that have an equivalent ground state for the above Hamiltonian (or, the exchange of FM and AFM bonds, which is another gauge choice).  Only when we explore perturbations to this Ising Hamiltonian in Sections \ref{Spet} and \ref{Sfield} does our particular gauge choice become important, and we will discuss it more there.

In this paper, we study the ground states of this model using an efficient Markov chain Monte Carlo (MCMC) method.  As described in Section \ref{CFA}, conventional local (or ``single-spin flip'' (SSF)) MCMC algorithms fail to explore the entire degenerate ground state ergodically, which leads to incorrect simulation results, in particular for perturbed models.  The problem can be alleviated with global loop and cluster algorithms \cite{BN_loop}, however those designed for use on corner-sharing triangular or tetrahedral lattices \cite{Isakov,JPC} do not generalize to the FF honeycomb Ising model.  Therefore, we develop a general {\it chain-flip} algorithm which allows for full ergodicity in the MCMC sampling of the ground-state manifold of the unperturbed model.  We demonstrate how this restored ergodicity uncovers a phase transition to a long-range ordered state in a perturbed model, whereas conventional algorithms with only local configuration changes do not.  

In Section \ref{Sfield} of this paper, we use our MCMC algorithm to explore the evolution of the ground state of the Hamiltonian in an applied external magnetic field, where we find two non-trivial higher-magnetization states with non-extensive entropy (scaling as $\sqrt{N}$).  In addition, for two critical field values bounding these states, we find special points of restored extensive entropy.
Remarkably, one is able to calculate the values of these ``reemergent'' extensive entropies exactly, and we find that they scale proportion to $N\ln \phi$, where $\phi$ is the golden ratio.  These values are confirmed by our MCMC simulations.

\section{The chain-flip algorithm} \label{CFA}

The most powerful method to study classically frustrated lattice spin models is MCMC.  The design of efficient algorithms has made much of the past discoveries in the field possible, but it is not without its difficulties.  In particular, the ability to study perturbed Hamiltonians, and phenomena related to the lifting of macroscopic Ising ground-state degeneracies, is critically dependent on global ``cluster'' algorithms which are able to traverse degenerate (or nearly degenerate) configurations in order to discover energetically preferred ground states.  In other words, local updates (like the SSF mentioned above) may suffer from a loss of {\it ergodicity}, or an exponential suppression of computational efficiency, effectively becoming frozen into particular states due to the presence of large energy barriers between configurations.  However, global updates may restore ergodicity to the algorithm.  This has been demonstrated with the development of loop algorithms in classical two-dimensional (2D) ice and vertex toy models \cite{BN_loop}, and has matured into a very general set of loop algorithms applicable to a wide range of models on corner-sharing triangular-based lattices in two dimensions (kagome) \cite{Isakov} and three dimensions (pyrochlore) \cite{JPC}.

\subsection{Description of the Algorithm}

In this section, we examine in more detail the mechanism by which SSF updates become inefficient in our FF honeycomb lattice Ising model Eq.~(\ref{FFh}), before developing a global ``chain-flip'' algorithm which restores ergodicity at low temperatures.  Consider the 
 classical SSF Metropolis algorithm, where configurational changes are made by attempting to flip each spin individually, finding the corresponding change in energy  $\Delta E=E_{\rm after} - E_{\rm before}$, and then accepting the flip with probability 
\begin{equation}
 P = \text{min}\left\{
 \begin{array}{rl} 
  &exp(\frac{-\Delta E}{T}) \\ 
  &1.
 \end{array}
 \right. \label{Boltzmann} 
\end{equation}
In this as in all frustrated models, the SSF method works well at higher temperatures $T$, but as the temperature becomes lower, the system begins to ``freeze'' (or lose ergodicity) into its disordered manifold of equal-energy states.
Each degenerate ground state configuration is at the bottom of a local energy well, which means that most ``nearby'' system configurations (configurations with only a few different spins) have a significantly higher energy.  As a result, the probability $P$ that any SSF which breaks out of the degenerate manifold is accepted becomes exponentially low.
To get from one ground state to another (i.e.~to move from one energy minimum to an adjacent one), multiple consecutive single spin flips are needed, and since the probability of any one spin being flipped is low, the chance that multiple spins are flipped consecutively is very small.  This means that the simulation dynamics become effectively frozen, or non-ergodic.  

In order to overcome this difficulty, one requires an algorithm that flips multiple spins simultaneously in a way that bypasses the large  energy barriers and tunnels from one ground state to the next.   We achieve this by introducing a ``chain-flip'' algorithm for the FF honeycomb Ising model.
To understand how this chain move works, it is first useful to understand the structure of ground states in the model  (see Fig.~\ref{latt_FFH} for one example).  In any given configuration, we call a bond unsatisfied if it has positive energy ($+J/4$), and satisfied otherwise ($-J/4$).  
The spin configurations that contribute to the ground state manifold are the ones in which every hexagon on the lattice has one and only one unsatisfied bond.
If flipping a group of spins creates as many unsatisfied bonds as it removes, 
such a process is equivalent to moving from one degenerate ground state to another.  This {\it net} zero-energy move is not possible with single-spin flips alone.

\begin{figure}
 \includegraphics[width=3in]{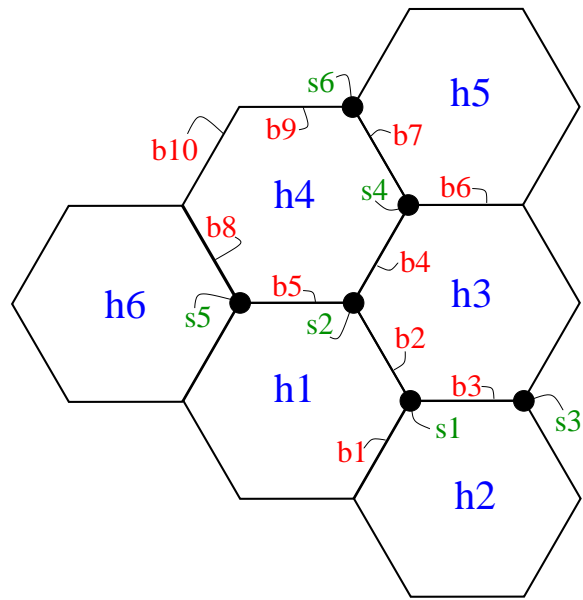}
 \caption{\label{ChainM}
(Color online) Labeled Lattice}
\end{figure}

Now we look at a method for finding a group of spins that creates as many unsatisfied bonds as it removes (Fig~\ref{ChainM}).  We build up a {\it chain} of spins by selecting vertices one at a time and counting the number of unsatisfied bonds that flipping the spin creates, versus how many 
it removes. 
Call the total number of unsatisfied bonds at any point in the algorithm the {\it net} unsatisfied bonds.
It is useful to note that flipping a spin changes the 
state of 
frustration of its three neighboring bonds.  We note that if at any point in building the chain we encounter a hexagon not initially in its ground state (i.e.~with more than one unsatisfied edge) we cancel the chain, thus ensuring the lattice remains locally in a ground state and ensuring detailed balance.

Our first step is to pick a hexagon at random and label its unsatisfied bond $b1$ (see Fig.~\ref{ChainM}).  Label the two hexagons that share that bond $h1$ and $h2$.  Then we randomly pick one of $b1$'s spins, and label it $s1$, and label the third hexagon that $s1$ is adjacent to $h3$.  Label the bond shared by $h1$ and $h3$ bond $b2$, and the bond shared by $h2$ and $h3$ bond $b3$.  Note that since $b2$ and $b3$ are edges of $h1$ and $h2$ respectively, neither of them can be unsatisfied.  Now store $s1$ as the first spin in our chain, and flip it.  Flipping $s1$ makes $b1$ satisfied and $b2$ and $b3$ unsatisfied, giving us $+1$ net unsatisfied bonds.

We know that $h3$ initially had one unsatisfied bond, and $h3$ has four bonds we have yet to consider: two adjacent to $b2$ and $b3$, and two opposite $b2$ and $b3$.  Label the spins that $s1$ shares with $b2$ and $b3$ with $s2$ and $s3$ respectively.  If $h3$'s initial unsatisfied bond is adjacent to $s2$, then $s2$ is now adjacent to two unsatisfied bonds (its third bond, being part of $h1$, is satisfied).  Thus we can flip $s2$ and create $-1$ unsatisfied bonds, giving our chain ($s1$ and $s2$) a total of $0$ net unsatisfied bonds when flipped, and we are done.  This argument works similarly if $h3$'s initial unsatisfied bond is adjacent to $s3$.  So we see that the minimum number of spins that need to be flipped to complete the chain algorithm is two.  Alternatively, the algorithm continues, if $h3$'s unsatisfied bond is opposite either $b2$ or $b3$.  Incidentally, the basic two-spin chain flip is analogous to the elementary dimer plaquette moved generated by the kinetic term in typical quantum dimer models \cite{RVB}.

To explore the more general case, assume without loss of generality that the unsatisfied bond occurs opposite to $b3$.  In this case, we want to flip $s2$, so label the second bond $s2$ shares with $h3$ bond $b4$, label $s2$'s third bond $b5$, and label $b4$'s other spin $s4$.  Bonds $b4$ and $b5$ are satisfied, so flipping $s2$ creates $+1$ unsatisfied bonds, for a total of $+2$ net unsatisfied bonds.  Now label $h3$'s unsatisfied bond $b6$, and $s4$'s third bond $b7$.  Hexagon $h3$ shares $b4$ with $h4$, and label the hexagon adjacent to both $h3$ and $h4$ hexagon $h5$.   Spin $s4$ has two neighboring unsatisfied bonds, $b6$ and the newly unsatisfied $b4$, and one neighboring satisfied bond, $b7$, which we know is satisfied because it is part of $h5$, which already has $b6$ as its unsatisfied bond.  Thus flipping $s4$ creates $-1$ unsatisfied bonds, for a total of $+1$ net unsatisfied bonds.  The chain must therefore continue; we now begin the recursive phase of the algorithm.

Hexagon $h4$ must initially have unsatisfied bonds, and so far we have looked at three of its bonds, $b5$, $b4$, and $b7$.  Bond $b5$ and $b7$ have been made unsatisfied by spin flips we've already done.  Label $b5$'s second spin $s5$ and $b7$'s second spin $s6$.  If $h4$'s initial unsatisfied bond is adjacent to $s5$ we can flip $s5$, creating $-1$ unsatisfied bonds, for a total of $0$ net unsatisfied bonds.  The algorithm is therefore done.  Similarly, if $h4$'s unsatisfied bond is adjacent to $s6$, one can flip $s6$ and the chain is done.  The recursive case comes when $h4$'s unsatisfied bond is opposite $b4$.  Label the bond adjacent to $s5$ on $h4$ bond $b8$, the bond adjacent to $s6$ on $h4$ hexagon $b9$, and the bond opposite $b4$ bond $b10$.  We now randomly choose either $b8$ or $b9$ and flip both of its spins.  Say without loss of generality that we chose $b8$.  Since $b5$ and $b10$ are unsatisfied and the other two bonds adjacent to $b8$ are satisfied, flipping both of $b8$'s spins results in $0$ unsatisfied edges, for a total of $+1$ net unsatisfied bonds.  Label the hexagon that shares $b8$ with $h4$ hexagon $h6$.

The recursion comes by realizing that our current state is the same as it was before we flipped $b8$'s two spins: $h6$ has one unsatisfied edge that we have not found yet, on one of three sides that we have not considered.  Furthermore, two of $h6$'s spins have been flipped.  Thus we can apply the same logic in choosing spins to add to the chain that we applied when choosing the last two.  Specifically, if $h6$'s unsatisfied bond is the one opposite $b8$, then we continue recursing.  If the unsatisfied bond is one of the other two possibilities, we can end the chain with $0$ net unsatisfied bonds, or in other words, with a chain that, when flipped, takes us to another ground state. 

In $T=0$ simulations of the Hamiltonian Eq.~(\ref{FFh}), where the model is expected to be in the ground-state, spins on completed chains can be flipped with probability unity.  However, in the case of perturbed Hamiltonians (see sections \ref{Spet} and \ref{Sfield}), the energy change $\Delta E$ incurred by the proposed chain-flip must be calculated.  Then, this proposed flip is accepted or rejected based on the Metropolis condition, Eq.~(\ref{Boltzmann}).

\subsection{Chain moves in the unperturbed model}

In order to test the efficiency of the chain move in a real simulation, in this section we present results for two MCMC codes for the unperturbed model Eq.~(\ref{FFh}), each employing a Metropolis algorithm with a Boltzmann probability Eq.~(\ref{Boltzmann}).   The first uses a conventional SSF Metropolis algorithm, the second a combination of SSF and chain-flip updates.  Specifically, the MCMC step using SSFs alone consists of attempting to flip each spin in the lattice {\it twice}. The MCMC step using chain moves consists of attempting SSF on each spin in the lattice {\it once}, followed by one attempted chain move for every $20$ spins in the lattice.  This convention for MCMC steps was chosen so that the CPU time of the chain-flip assisted step is roughly equal to that of the step using SSFs alone, thereby allowing us to directly compare results without relying on formal autocorrelation measurements.  We note of course that, in a working MCMC code, other conventions for the Monte Carlo ``step'' may be chosen by the practitioner.    

In this work, simulations were performed at finite temperatures and on finite system sizes ranging from a few hundred to tens of thousands of spins, using of order $10^6$ MCMC steps.  From such simulations, we examine the impact of the chain move certain thermodynamic quantities, in particular, the energy $E$, specific heat $C ={\partial E}/{\partial T}$, and magnetization per spin $M =1/N\sum_i S^z_i$.  Further, we restrict ourselves to looking at two procedures for obtaining such finite-$T$ data: 1) an {\it annealing} (or slow-cooling) algorithm, and 2) a {\it quenched} (or rapid-cooling) algorithm, details of which are reported below.

\begin{figure}
 \includegraphics[width=3.2in]{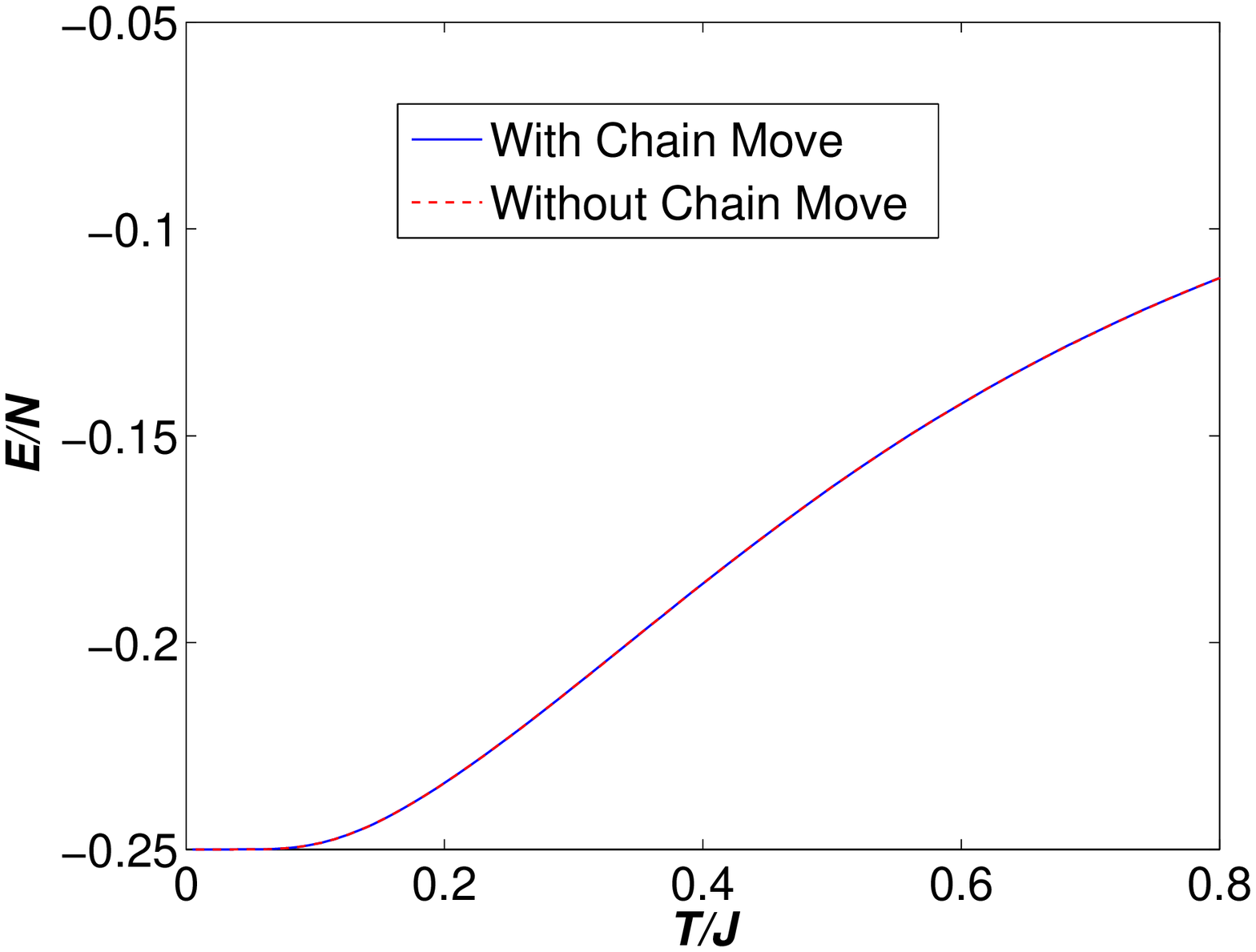}
  \includegraphics[width=3.2in]{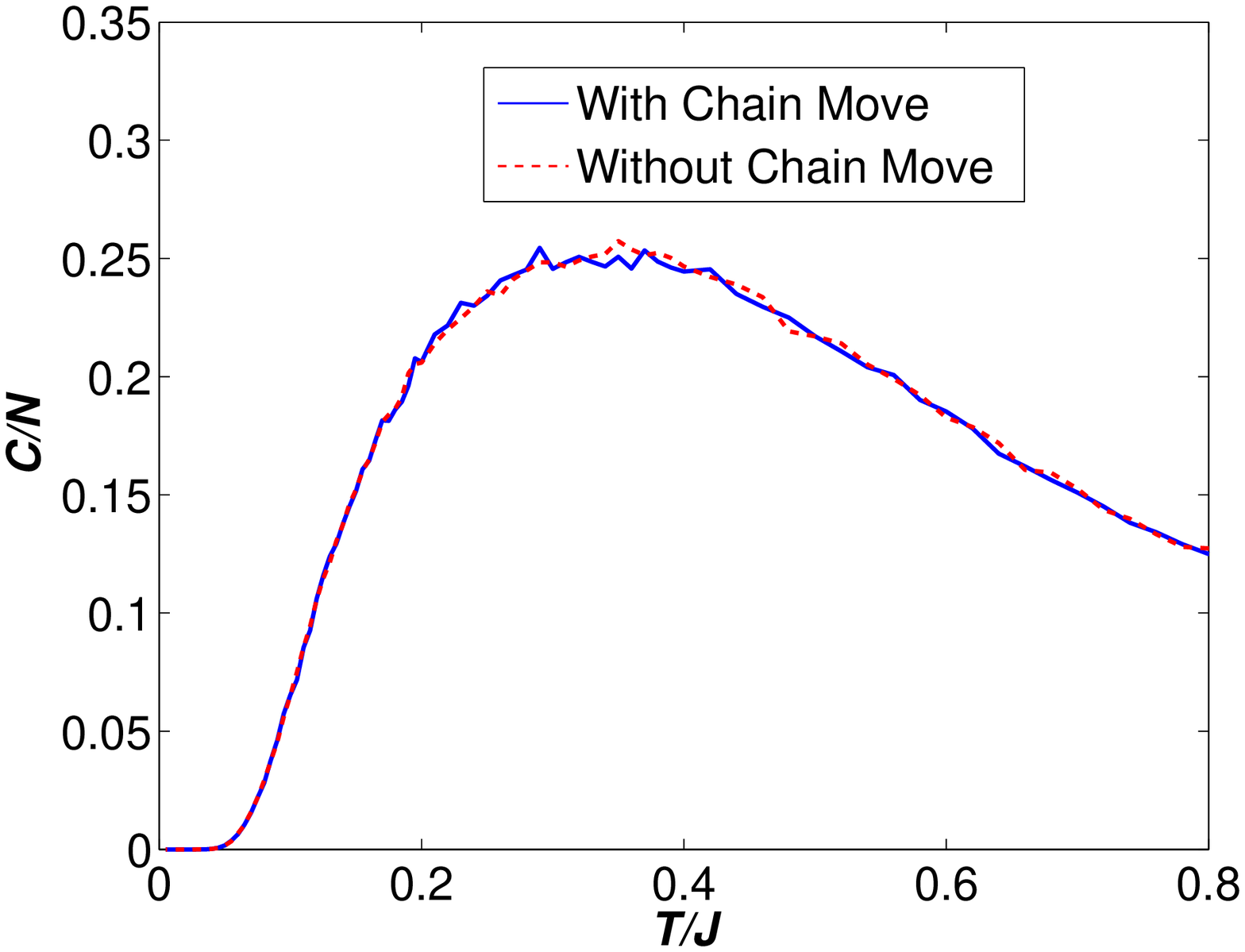}
  \includegraphics[width=3.3in]{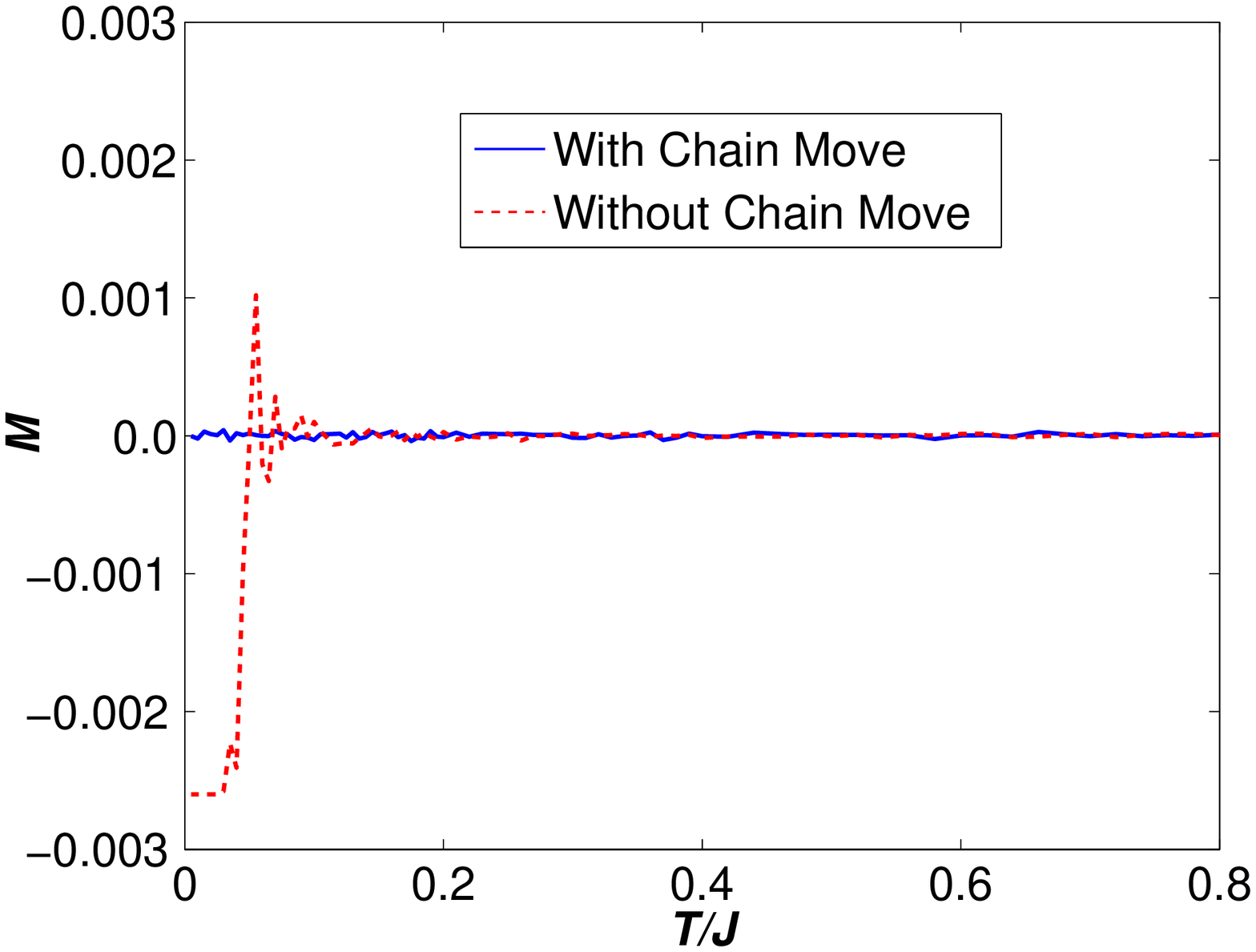}
  \caption{ \label{ECunp} (Color online) The energy, specific heat, and magnetization of a simulation of $20000$ Ising spins on the unperturbed FF honeycomb model.
  }
\end{figure}

First, we examine data obtained through annealing simulations.
An annealing procedure is often employed in simulations of models with long time-scales or glassy dynamics, and is known to help reach even complex ground states using very simple local (SSF) algorithms.  In an annealing run, the simulation is started at a high temperature $T/J>>1$, and the usual MCMC algorithm (a series of equilibriation and production steps) is employed.  After sufficient data is gathered, the temperature is lowered by a small step, {\it keeping} the system configuration from the previous (higher-temperature) step. The MCMC algorithm is then repeated, and the temperature is lowered again until the system settles in its ground state.

Figure~\ref{ECunp} illustrates the results for the energy $E$, the specific heat $C$, and the magnetization $M$ of the annealing algorithm for a system of $20000$ Ising spins in the Ising model.  It is clear that the results for $E$ and $C$ are similar for both types of MCMC step, and that both realize the proper ground state of the model, with energy per spin ${E}/{N} = -{J}/{4}$.  Further, integration of ${C}/{T}$ (see e.g.~Ref.~\cite{JPC}) for both of these simulation runs reveals that the model retains a residual entropy in its ground state of ${S}/{N} = 0.214$ to within numerical (1\%) accuracy \cite{Wolff}.  The lack of difference between the SSF and the chain algorithm data can be explained largely as the success of annealing: even without the chain move, annealing allows the SSF algorithm alone to find a ground state.  However, the single spin flips cannot move {\it between} degenerate ground states at very low $T$;  $E$ and $C$ are simply unaffected as every ground state has the same energy.  This is not the case with the magnetization $M$, as seen in Figure~\ref{ECunp}.  Clearly, the expectation that the ground state magnetization per spin should be tightly distributed around a mean of $M=0$ can be violated in the SSF algorithm, where degenerate configurations with higher magnetization can be frozen in, as illustrated.  However, with the chain-flip algorithm, the expected convergence to $M=0$ is found with high accuracy.

 \begin{figure}
 \includegraphics[width=2.8in]{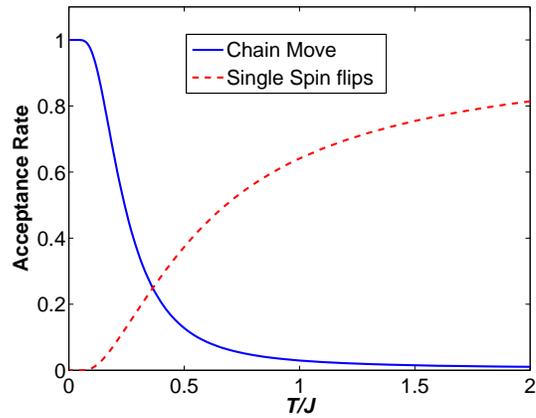}
  \caption{ \label{AccRate} (Color online) The acceptance rate of single spin flips and chain moves across a range of temperatures.
  }
\end{figure}

Our next observation is of the acceptance rate of the chain moves (Fig.~\ref{AccRate}).  The single spin flips work well until lower temperatures are reached, at which point their acceptance rate drops off to zero.  Comparing to Fig.~\ref{ECunp}, this happens at a temperature where the system has realized the degenerate ground state. 
The chain moves, in contrast, have a very low acceptance rate at high temperatures.  This is due to the large number of chains being aborted during generation, when the construction algorithm encounters a hexagon with the ``incorrect'' number of unsatisfied bonds (i.e.~not one).  
  As the temperature lowers and the system reaches the ground state manifold, the chain-move acceptance rate climbs to unity.  Clearly, a combination of SSF and chain-flips is needed in order to give a MCMC step with a reasonable acceptance rate across all temperatures.

\begin{figure}
 \includegraphics[width=3.2in]{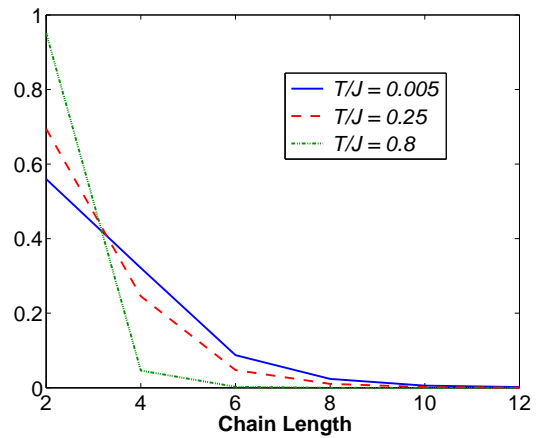}
 \caption{\label{Chain_Dist}
(Color online) The distribution of chain lengths on a lattice of $20000$ spins.
 }
\end{figure}
It is interesting to consider the size of chains produced by the algorithm.  Note first that every chain must have an even size to ensure an even number of ``boundary'' bonds (see Fig.~\ref{ChainM}).  In Fig.~\ref{Chain_Dist}, we see that in a histogram of data collected for several parameter values, the most common chain size is two, with the distribution of chains decreasing rapidly with their size.  
At high temperatures this distribution is weighted heavily towards chains of size two, but as temperature decreases the distribution flattens out somewhat, making longer chains more probable.  
We also note that the distribution of chain sizes is essentially unaffected by lattice size, since apparently very few chains are long enough to act on more than a very local area of the lattice.

We complete our examination of the two possible MCMC algorithms mentioned above, turning now to a discussion of the {\it quenching} algorithm.  Quenching refers to the procedure whereby a simulation is run completely at a single (usually low) temperature, in an attempt to obtain the ground state quickly without annealing.  The desire to use a quenching algorithm is obvious if one is interested only in the low-temperature properties of the model, which is often the case, and it is widely used since it is also the easier of the two algorithms to implement.  However, without the history of higher-temperature configurations provided by an annealing procedure, it is often observed that simulations at low temperature have a more difficult time settling into their true ground state configuration, as it is possible to get trapped in local energy minima separated by large energy barriers that single-spin flips have difficulty overcoming.

\begin{figure}
 \includegraphics[width=3.2in]{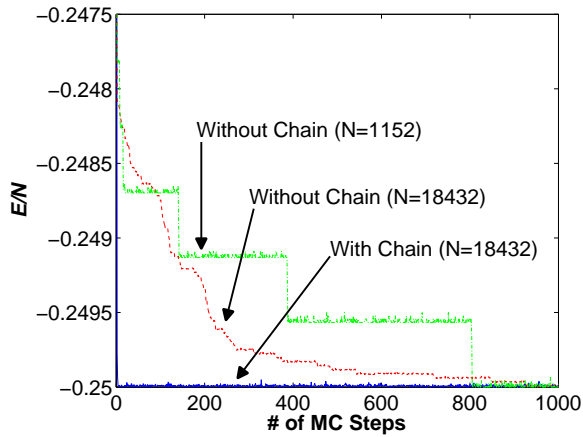}
 \caption{\label{quench}
(Color online) The energy of a simulation of $20000$ Ising spin in the unperturbed model at ${T}/{J}=0.05$, as a function of Monte Carlo (MC) step.
 }
\end{figure}

Figure~\ref{quench} illustrates the energy of three different quenched simulations, all begun in a random initial state (at step zero), as a function of the number of MCMC steps.  
It is clear that the chain-flip simulation reaches the proper ground state energy in a fraction of the number of steps that the simulations without it take.  The effects of the energy barriers and local minima on simulations using only single spin flips is most obvious in the plot of the energy of the smaller lattice of size $N = 1152$.  The plateaus correspond to local minima in the energy, and the jumps between plateaus correspond to several spins being flipped consecutively, allowing the simulation to ``bypass''  the energy barrier.  The simulation using chain-flips is virtually unaffected by the energy barriers, and finds the ground state in less than $50$ MCMC steps.  This figure is therefore a testament to the increased efficiency of the algorithm with the chain move in reaching the ground state.  

In this section, we have demonstrated that MCMC simulations employing the chain-flip algorithm can both realize the ground state of the unperturbed FF honeycomb lattice Ising model more efficiently that those employing single spin flips alone, and also remain unfrozen once in this ground state.  We now demonstrate the usefulness of the chain move in simulations where the Hamiltonian has been perturbed by a small interaction that lifts the degeneracy of the ground state manifold.

\subsection{Chain moves in a perturbed model} \label{Spet}

In physical cases of interest, for example in the modeling of real materials, one typically expects a more complicated Hamiltonian than Eq.~(\ref{FFh}), often taking the form of small perturbations added to (or modifying) the simple Ising interaction.  In many applications, these perturbations require no modification of existing MCMC schemes.  However in the case where the unperturbed ground state is an extensively degenerate manifold of equal-energy states (such as is the case with the FF honeycomb Ising model), this is not true.  Specifically, if small perturbative interactions lift the degeneracy of the manifold by energetically favoring one or more specific configurations (e.g.~promoting long-range order), it has been demonstrated that SSFs in a MCMC scheme can fail to find this true ground state \cite{JPC}.   

In our FF honeycomb model, one may see the dynamical freezing of the SSFs by inspecting the acceptance rate in Fig.~\ref{AccRate}, which also suggests that the chain-flips successfully explore the degenerate manifold of states at very low temperatures.
We test this idea by introducing a small perturbation in the Hamiltonian which slightly {\it weakens} the ferromagnetic bonds in Fig.~\ref{latt_FFH}.  The Hamiltonian can be written as 
\begin{equation}
H =  J\sum_{\langle ij \rangle}  S^z_i S^z_j  \delta_{\langle i j \rangle, a}- J' \sum_{\langle ij \rangle}  S^z_i S^z_j  \delta_{\langle i j \rangle, b} \label{Pertb}
\end{equation}
where the first term is the Hamiltonian for the antiferromagnetic bonds, and the second for the ferromagnetic bonds (i.e.~both $J$ and $J'$ are positive constants -- see Fig.~\ref{latt_FFH} for the bond labels $a$ and $b$).  In the following discussion, we set $J'/J=0.90$.  With this slight perturbation, we expect that the system will select a unique ground state with long-range order from the extensive manifold of states in the unperturbed model.  Since the ferromagnetic bond is slightly weakened, this order will be one where the unsatisfied bond (one per hexagon) is placed uniquely on the ferromagnetic bond $b$.  We expect the energy per spin of this ground state to be $-{41}J/{160}$, which is less than the $-{J}/{4}$ of the unperturbed model.
\begin{figure}
 \includegraphics[width=3.2in]{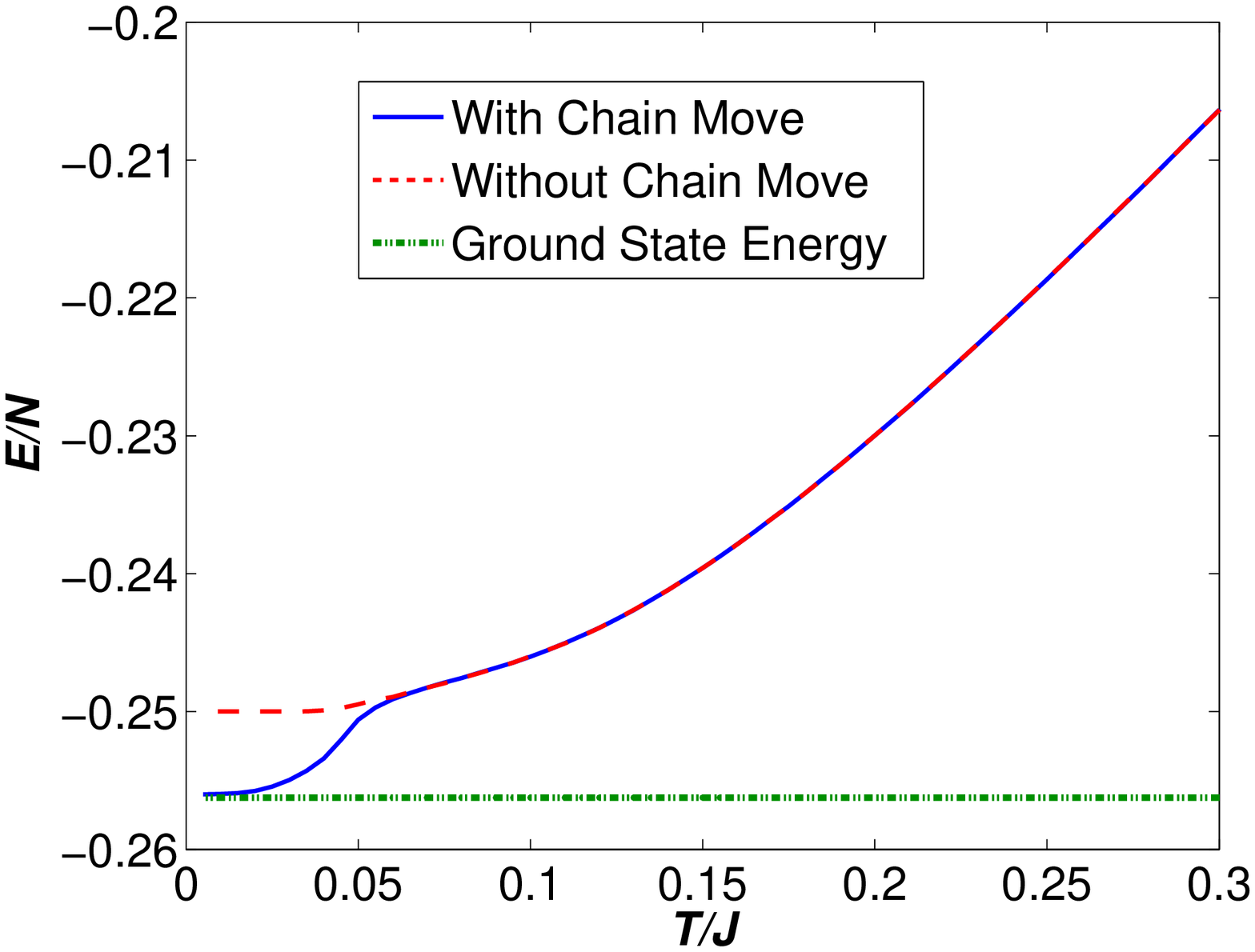}
  \includegraphics[width=3.2in]{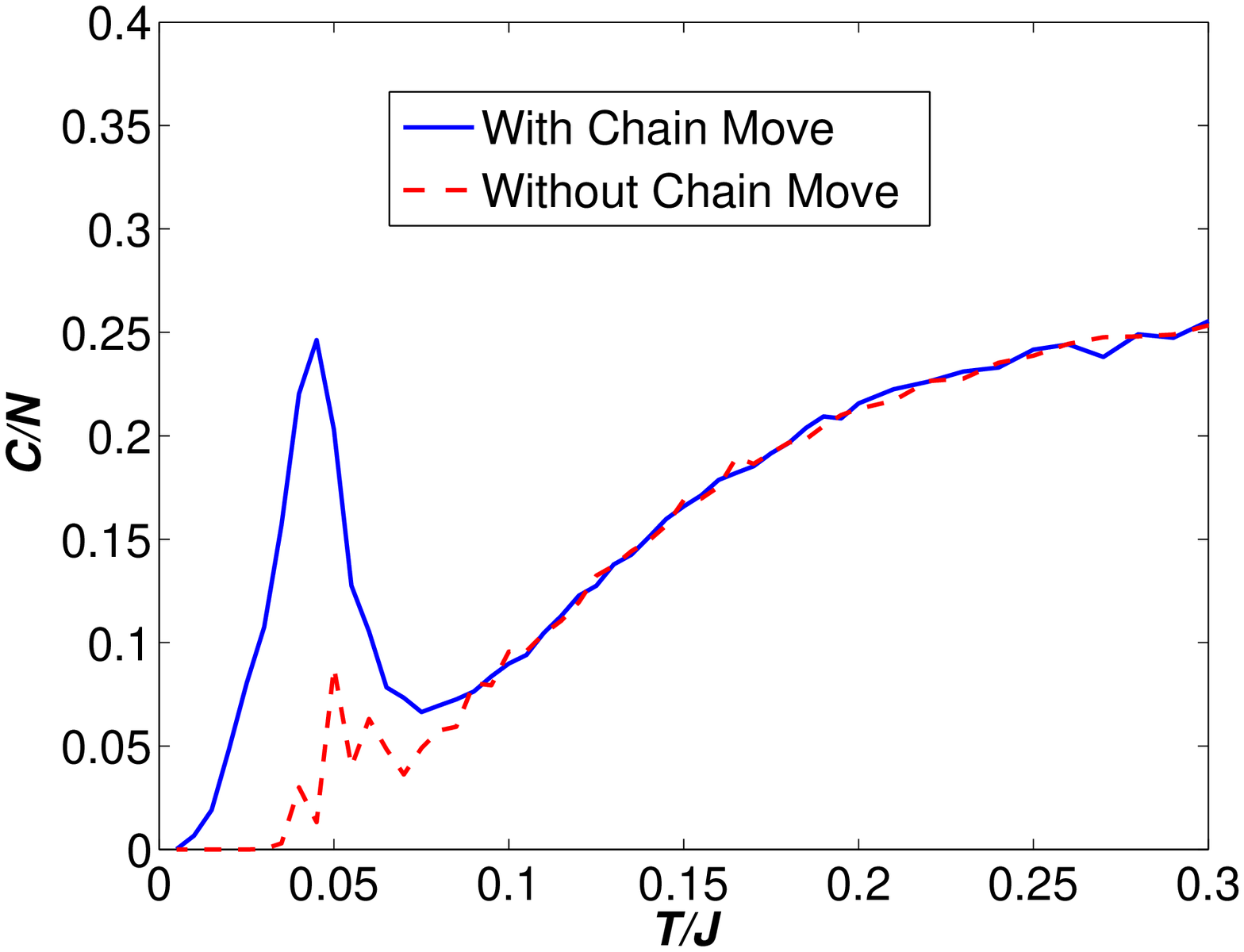}
  \caption{ \label{ECpet} (Color online) The energy and specific heat of a simulation of 20000 Ising spins on the perturbed FF honeycomb model, Eq.~(\ref{Pertb}).
  }
\end{figure}

Indeed Fig.~\ref{ECpet} shows that the chain moves significantly modify the behavior of the MCMC simulation.  Results presented there are for an annealing algorithm, and in contrast to the unperturbed case, it is clear that the MCMC using SSFs does not find the correct ground state, while the MCMC using chain-flips does.  In addition, the specific heat curve in Fig.~\ref{ECpet} shows a large peak in the algorithm using the chain move that is {\it not} present above the noise seen in the algorithm not using the chain move.   One observes a dynamical freezing of the spin configuration in the SSF algorithm, where the MCMC simulation no longer is able to sample low-lying states, and hence eventually freezes into a disordered state.  In contrast, the chain algorithm is able to find a phase transition to a long-range ordered state, promoted by the perturbed Hamiltonian, as evident from the peak in $C$.  Integration of this specific heat peak (over $T$)
 in Fig.~\ref{ECpet} with the chain moves finds all of the expected $\ln(2)$ entropy, confirming the development of a unique groundstate at $T=0$.  In contrast, with the SSF algorithm only, the the full $\ln(2)$ entropy is not recovered by the integration, indicating that  a long-range ordered ground state is not found.

\section{Groundstates in an external magnetic field} \label{Sfield}

We now turn to a consideration of the fully-frustrated honeycomb lattice Ising model, with the physically important perturbation of an external magnetic field:
\begin{equation}
H =  J\sum_{\langle ij \rangle} (-1)^{\delta_{\langle i j \rangle, b}} S^z_i S^z_j + h \sum_i S^z_i \label{FFh_hz}.
\end{equation}
The inclusion of a symmetry-breaking magnetic field is known to lift the degeneracy of the ground-state manifold in some models, e.g.~the frustrated Ising AFM on a triangular lattice \cite{Murthy}.  However, in other cases, such as the frustrated Ising AFM on the kagome lattice, this is not the case, and the presence of a perturbative external magnetic field reduces the degeneracy but does not lift it all together \cite{Moessner1}.  

\begin{figure}
 \includegraphics[width=3.2in]{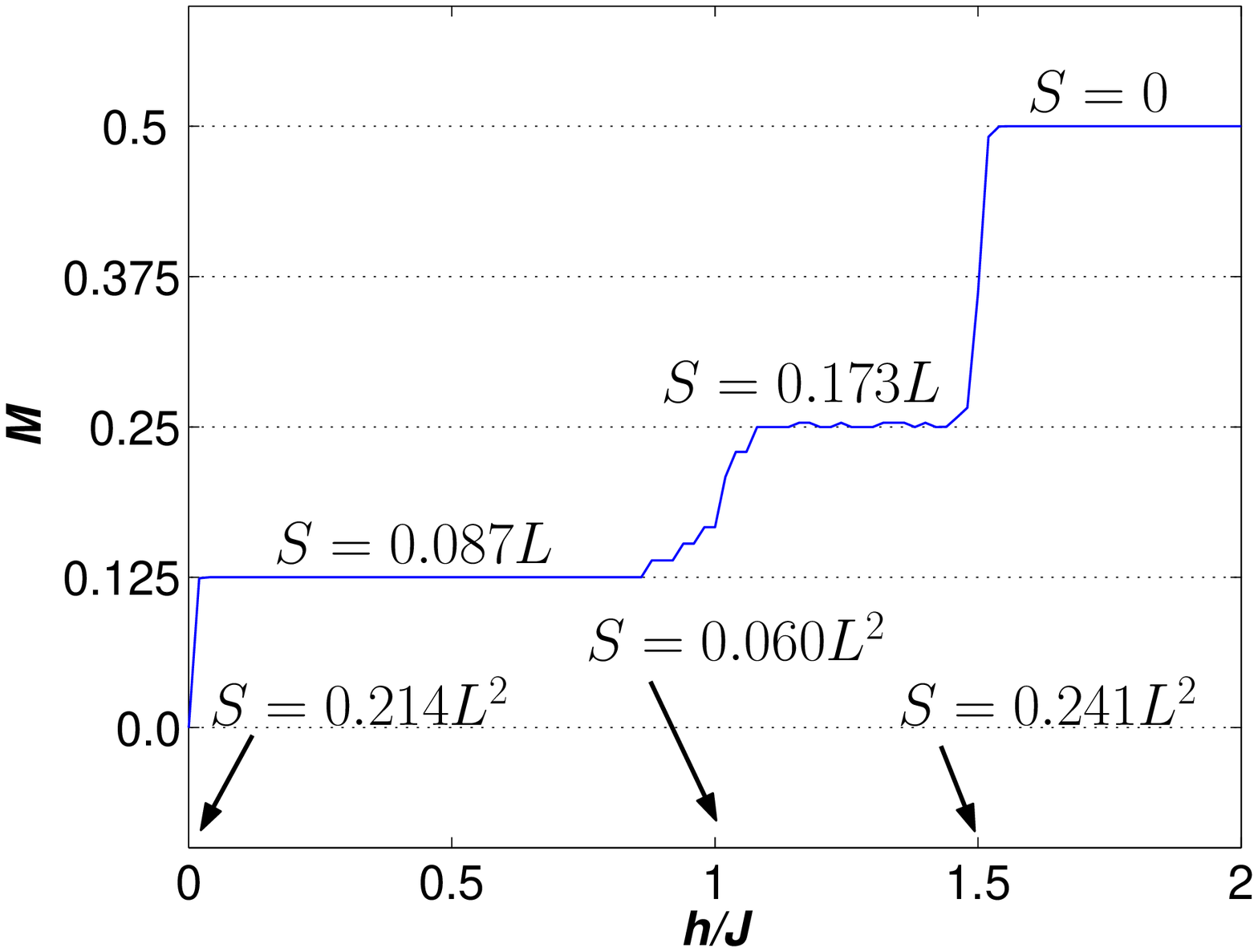}
 \caption{\label{hfield}
(Color online) The magnetization per spin of a simulation of 288 Ising spins ($L=12$) in the model Eq.~(\ref{FFh_hz}) at ${T}/{J}=0.005$.  The plateaus at $M=1/8$ and $M=1/4$ correspond to partially-ordered states, described in the text.  The (asymptotic) values of the residual $T=0$ entropy $S$ of the ground state configurations are also labelled.  The extensive entropy spikes which occur a the special values of $h/J=1$ and $h/J=3/2$ are discussed in Section~\ref{exactS}.}
\end{figure}

\subsection{Magnetization plateaus and entropy spikes } \label{Spike}

Using MCMC simulations with the SSF and chain-flip algorithm, we study the ground state of the Hamiltonian, Eq.~(\ref{FFh_hz}), on the honeycomb lattice.  Figure~\ref{hfield} shows the evolution of the magnetization per spin $M$ as a function of the applied field.  Immediately upon application, the field promotes the development of a plateau with magnetization $M={1}/{8}$.  Inspection of simulation configurations on this plateau reveal that it corresponds to a partial ordering of spins.  

\begin{figure}[ht]
 \includegraphics[width=3in]{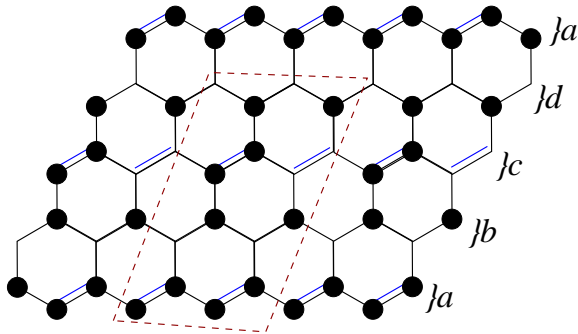}
 \caption{\label{Field} (Color online)  The ground-state spin configuration of the Hamiltonian Eq.~(\ref{FFh_hz}) at magnetization $M=1/8$ ($0<h/J<1$).  Black dots represent $S^z$ spin-up, and empty sites spin-down.  The 16 sites making up the unit cell are outlined by the dashed (red) line.  To obtain the $M=1/4$ state, the down-spins on the zigzag ``row'' labelled $c$ must be flipped to up-spins.
 }  
\end{figure}

This order is illustrated in Fig.~\ref{Field}; it can be described in terms of horizontal zig-zag ``rows'' (labelled $a$ to $d$ in Fig.~\ref{Field}).   There, every {\it second} row with ferromagnetic bonds (labelled $a$) is fully polarized (all spin-up).  This choice of ordering pattern is obviously not gauge-invariant, since the pattern of FM bonds has been chosen to break the lattice rotational symmetry.  It can be seen however that it lowers the energy of the spins associated with these bonds in the magnetic field, while retaining a configuration that is a member of the ground state manifold in the unperturbed model (i.e.~one and only one unsatisfied bond per hexagon).  The rows adjacent to the fully polarized row, (i.e.~every row without a ferromagnetic bond, or rows $b$ and $d$ in Fig~\ref{Field}), are forced into a specific configuration, with alternating up and down spins, in order to maintain one unsatisfied bond per hexagon.  Finally, the remaining rows with ferromagnetic bonds (the $c$ rows in Fig~\ref{Field}) each have two possible configurations, each alternating two up spins, with two down spins along their ferromagnetic bonds.  Remarkably, this is {\it not} a unique long-range ordered state in two dimensions, since the FM rows of spins $c$ sandwiched in between the fully polarized FM rows $a$ are ordered {\it independent} of the FM $c$ rows above and below.  Thus, a ground state entropy remains.  However, it is no longer extensive (i.e.~scaling as $L^2$), but scales as $L$, the lattice linear dimension. See Section \ref{exactS} below for an exact expression for this entropy.

The $M={1}/{8}$ semi-ordered state remains stable to moderately large applied fields (as evident by the plateau in Fig.~\ref{hfield}), until for $h/J>1$ a second plateau is reached at $M={1}/{4}$.  We note that for this plateau, the ground state is forced out of the degenerate manifold of states (with one and only one unsatisfied bond per hexagon).  Hence, chain moves cease to be effective (although SSFs still contribute), resulting in the increased noise in the magnetization plateau in Fig.~\ref{hfield} and a rounding of the associated transition.  From observation of simulation configurations, it is apparent that this plateau corresponds to the flipping of all down-spins to up-spins on row $c$ (Fig.~\ref{Field}).  This configuration is again not truly long-ranged ordered in two dimensions.  Rather, in this case the antiferromagnetic rows $b$ and $d$ have become independent of each other, and the ferromagnetic row $c$ has become fixed.  Thus, we find that the entropy of the ${1}/{4}$ plateau is actually twice as large as the entropy of the ${1}/{8}$ plateau, although still scaling as the linear system size $L$ (see Section \ref{exactS}).
Finally, for applied magnetic field $h/J>{3}/{2}$, the remaining down-spins on the $b$ and $d$ rows flip to up-spins, and the system becomes fully-polarized.


A curious phenomenon occurs in the transition regions between the three magnetization plateaus illustrated in Fig.~\ref{hfield}.  For the special values $h/J=1$ and $h/J=3/2$, the model transitions from a configuration of decoupled quasi-1D ordered chains to once again being a disordered 2D system.  
In other words, at precisely these critical values of $h/J$, the model becomes ``accidentally'' macroscopically degenerate due to fine-tuning of the magnetic field.  At $h/J=1$, this degeneracy occurs between states like in Fig.~\ref{Field}, and states where {\it all} spins on all zig-zag row $c$ are up.  Similarly, the degeneracy at $h/J=3/2$ occurs between states like this last state, and the $M=1$ state where the remaining spins (on the zig-zag rows $b$ and $d$) flip up.  For these two special field values where the fine-tuning of $h/J$ causes an accidental degeneracy, we expect a reemergence of a residual ground state entropy, scaling as the system size $N$ (similar to the $h=0$ case).
Remarkably, one is able to calculate the values of these reemergent extensive entropies exactly in this model.  This is shown in the next section, where we discover that the asymptotic residual entropies are
\begin{eqnarray}
S/N&=& \ln \phi / 8\approx0.06015 \hspace{.5cm} {\rm for} \hspace{3mm} h/J=1, \\
S/N&=& \ln \phi / 2\approx0.24061 \hspace{.5cm} {\rm for} \hspace{3mm} h/J=3/2.
\end{eqnarray}
Here, $\phi \approx 1.618$ is the golden ratio.
Numerically, one can measure the values of the residual entropy via our MCMC simulations as the difference of the integral of $C/T$ (over all $T$) from $\ln(2)$.  For a lattice of size $L=32$, we obtain
$S=0.0604(2)$ at $h/J=1$, and $S=0.2398(2)$ at $h/J=3/2$,
where finite-size trends clearly suggest that the MCMC approaches the above asymptotic results in the thermodynamic limit, to within error bars.
The derivation of the exact asymptotic results is presented in detail in the next section.

\subsection{Exact entropy calculations} \label{exactS}
In this section we derive exact expressions for the finite-size entropy of the system with external magnetic field for varying $h/J$.  
In our derivation, we concentrate on linear lattice sizes $L$ that are a multiple of 4, based on the assumption (e.g.~from Fig.~\ref{Field} and discussions in Ref.~\cite{Moessner1}) that the smallest unit cell that contains the ordered or partially-ordered structure is commensurate with $L=4$.
 We also give approximate expressions for large lattice size $L$, and asymptotic values for $L$ tending to infinity.
In our presentation, we refer to Fig.\ \ref{Field}. 
\subsubsection{$0<h/J<1$}
As discussed above, for $0<h/J<1$, ground states are such that rows $c$ can each have two configurations (with pairs of spins alternating up and down), and there are $L/4$ rows $c$. Rows $a$ have all spins up, and rows $b$ and $d$ are fixed as in Fig.\ \ref{Field}. This gives $2^{L/4}$ configurations, but the roles of rows $a$ and $c$ can be switched, which leads to an additional doubling of this number of configurations. In this way we obtain
\begin{align}
\Omega&=2^{L/4+1},\\
S&=\ln{\Omega}=\left({L\over 4}+1\right)\ln{2},\\
{S\over N}&=\left({1\over 8 L}+{1\over 2 L^2}\right)\ln{2},
\end{align}
giving $S \sim  \ln(2)/8 \cdot L \approx 0.087L$ in the limit of large $L$.
\subsubsection{$1<h/J<3/2$}
We now extend our notation and call rows $a$ and rows $c$ rows of type $A$ ($A=\{a,c\}$) and, similarly, $B=\{b,d\}$.
As discussed above, for $1<h/J<3/2$, ground states are such that each row $B$ has its spins alternating up and down, and is thus in one of two possible states, independently from the other rows $B$: either all its up spins are adjacent to the row $A$ above it, or to the row $A$ below it. For each row $B$, let's assign a binary digit 1 to the former case (spins up are adjacent to the row $A$ above the row $B$), and 0 in the other case. Let $m=L/2$ be the number of rows $B$ on the lattice. Then there are $2^m$ configurations for the rows $B$, and we get
\begin{align}
\Omega&=2^{L/2},\\
S&=\ln{\Omega}={L\over 2}\ln{2},\\
{S\over N}&={1\over 4 L}\ln{2},
\end{align}
i.e.~$S \approx 0.173L$ in the limit of large $L$.
\subsubsection{$h/J>3/2$}
For $h/J>3/2$ there is a single ground state, with all spins up, and we get
\begin{align}
\Omega&=1,\\
S&=\ln{\Omega}=0,\\
{S\over N}&=0.
\end{align}

\begin{figure}
 \includegraphics[width=3in]{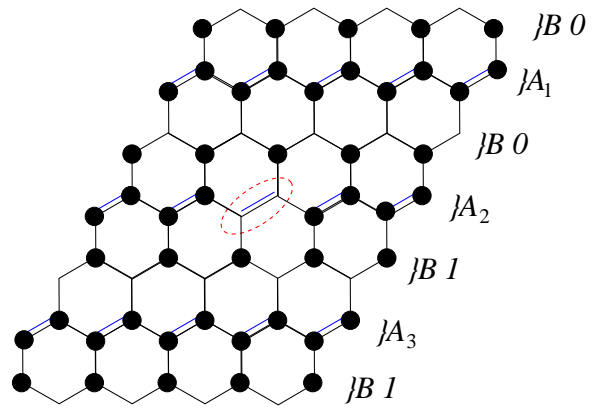}
 \caption{\label{hj1Field} (Color online)  The ground state spin configuration for $h/J=1$.  The binary digits 1 or 0 are associated with rows of type $B$.  Pairs of spins on rows of type $A$ can be flipped if that row is ``above'' a $B$ row of type 1, and ``below'' a $B$ row of type 0, for example the spins circled by the dashed (red) line in row $A_2$.  Spins on rows $A_1$ and $A_3$ may not be flipped.
 }  
\end{figure}
\subsubsection{$h/J=1$}

We first observe that all ground states for $1<h/J<3/2$ are still ground states for $h/J=1$. (Recall that these states have all spins up on rows $A$, and spins alternating up and down on rows $B$). However, there are now many additional ground states, because, for every of the $2^m$ configurations of rows $B$, some of the rows $A$ are allowed to flip some of their pairs of spins that are connected by a ferromagnetic bond from up to down. Indeed, rows $A$ that have their upper neighbor row $B$ in the 0 state, and their lower neighbor row $B$ in the 1 state, are allowed to flip pairs of spins without changing the energy, as long as no two adjacent pairs have spin down (see Fig.~\ref{hj1Field}). Note that this corresponds to the case that, for a pair of up spins in rows $A$, all neighboring spins are up as well.
For these rows $A$, flipping a spin pair (which has neighboring pairs up) from up to down results in a magnetic energy change of $4\, h/2$ (two spins flipped from up to down), and the energy change from the bonds is $-4\, J/2$, because four bonds become satisfied. At $h/J=1$ the energy thus remains unchanged. We now want to count how many states can be obtained in this way.

To this end, we first investigate how many rows $A$ are allowed to flip pairs of spins, for a given configuration of rows $B$. Every configuration of rows $B$ can be represented by an $m$-digit binary number (see above), and every transition from a 0 to a 1 in this binary number corresponds to a row $A$ whose spin pairs can be flipped. (Note that we have to include the periodic case, where the last digit is 0 and the first digit is 1, in our count.) Let $R(m,d)$ be the number of $m$-digit binary numbers with $d$ transitions from 0 to 1. It is shown in Appendix A that
\begin{equation}
R(m,d)=2\left(\begin{array}{c} m\\2\, d\end{array}\right).
\label{eq:Rmd}
\end{equation}

Next we have to count, for each row $A$ that is allowed to flip its pairs, how many configurations there are in which no two adjacent pairs are down. Let $n=L$ be the number of pairs on a row $A$, and let $g(n)$ be the number of configurations in which no two adjacent pairs are down. (Again, we have to include the periodic case in our count.) Appendix B shows that
$g(n)$ satisfies the Fibonacci recurrence equation, with exact solution
\begin{equation}
g(n)=c_1\,\phi^n+c_2\,\theta^n,
\label{eq:FibSol}
\end{equation}
where
\begin{align}
\phi&={1+\sqrt{5} \over 2}\approx1.6180,\\
\theta&={1-\sqrt{5} \over 2}\approx-0.6180,\\
c_1&={5+3\,\sqrt{5}\over 10}\approx1.1708,\\
c_2&={5-3\,\sqrt{5}\over 10}\approx-0.1708.
\end{align}
Note that, for large $L$, $g(L)$ can be approximated well by
\begin{equation}
\hat{g}(L)=c_1\,\phi^L.
\end{equation}

The number of ground states, $\Omega$, is now given by
\begin{equation}
\Omega=\sum_{d=0}^{L/4} R(L/2,d)\,g(L)^d.
\label{eq:sum}
\end{equation}
(Note that there are at most $d=L/4$ transitions from 0 to 1 in a binary number with $m=L/2$ digits.)

It is shown in Appendix C that a closed-form expression for this sum is given by
\begin{equation}
\Omega=\left(1+\sqrt{g(L)}\right)^{L/2}+\left(1-\sqrt{g(L)}\right)^{L/2},
\label{eq:omega}
\end{equation}
and then $S=\ln{\Omega}$ and $S/N=\ln{\Omega}/(2L^2)$.

For large $L$, we can approximate Eq.\ (\ref{eq:omega}) as
\begin{equation}
\hat{\Omega}=2\,\hat{g}(L)^{L/4}=2\,c_1^{L/4}\,\phi^{L^2/4},
\label{eq:omegaHat}
\end{equation}
which leads to
\begin{align}
\hat{S}&=\ln{\hat{\Omega}}=\ln{2}+{L\over 4}\,\ln{c_1}+{L^2\over 4}\,\ln{\phi},\\
{\hat{S}\over N}&={\ln{2}\over 2\,L^2}+{1\over 8\,L}\,\ln{c_1}+{1\over8}\,\ln{\phi}. \label{Ehj1}
\end{align}
We thus find an asymptotic value for $S/N$ equal to $\ln{\phi}/8\approx0.06015$.

\begin{figure}
 \includegraphics[width=3in]{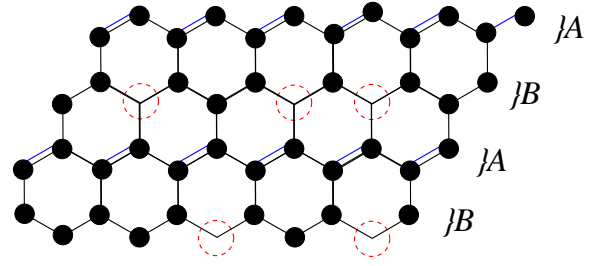}
 \caption{\label{hj32Field} (Color online)  The ground state spin configuration for $h/J=3/2$.  Spins on rows of type $B$ can be flipped as long as the condition of not having two adjacent down spins is satisfied.  Down spins circled by a dashed (red) line can be flipped to up spins without a change in energy.
 }  
\end{figure}
\subsubsection{$h/J=3/2$} \label{Ent32}
We saw above that, for $1<h/J<3/2$, ground states have all spins up in all rows $A$, and spins alternating up and down in rows $B$. These states are all ground states at $h/J=3/2$ as well, but there are many additional ground states, because rows $B$ can now flip some of their spins up or down without changing the energy, as long as no two adjacent spins are down (see Fig.~\ref{hj32Field}).
Indeed, flipping a spin (which has its three neighboring spins up) from up to down results in a magnetic energy change of $2\, h/2$ (one spin flipped from up to down), and the energy change from the bonds is $-3\, J/2$, because three bonds become satisfied. At $h/J=3/2$ the energy thus remains unchanged. We now want to count how many states can be obtained in this way.

First, there are $m=L/2$ rows B that can change some of their spins independently.
Second, every row $B$ has $n=2\,L$ spins that can be flipped up our down as long as no two adjacent spins are down. The number of valid configurations for each row $B$ is thus given by $g(2\,L)$, with $g(n)$ the specific solution of the Fibonacci equation given in Eq.\ (\ref{eq:FibSol}). This gives
\begin{equation}
\Omega=g(n)^{m}=g(2\,L)^{L/2},
\label{eq:omega2}
\end{equation}
and then $S=\ln{\Omega}$ and $S/N=\ln{\Omega}/(2L^2)$.

For large $L$, we can approximate Eq.\ (\ref{eq:omega2}) as
\begin{equation}
\hat{\Omega}=\hat{g}(2\,L)^{L/2}=c_1^{L/2}\,\phi^{L^2},
\label{eq:omegaHat2}
\end{equation}
which leads to
\begin{align}
\hat{S}&=\ln{\hat{\Omega}}={L\over 2}\,\ln{c_1}+{L^2}\,\ln{\phi},\\
{\hat{S}\over N}&={1\over 4\,L}\,\ln{c_1}+{1\over 2}\,\ln{\phi}. \label{Ehj32}
\end{align}
We thus find an asymptotic value for $S/N$ equal to $\ln{\phi}/2\approx0.24061$.


\section{Discussion}

In this paper, we have developed a global chain-flip algorithm for Markov Chain Monte Carlo simulations of the fully frustrated honeycomb lattice Ising model.  Chain-flips are used to complement conventional single-spin flips, in parameter regimes where the MCMC simulation is required to explore the model's extensively-degenerate ground state manifold of minimally-frustrated spin configurations.  
We have demonstrated, through careful numerical simulations, that the chain-flip algorithm both increases simulation efficiency, and restores ergodicity in the sampling of the degenerate manifold of states.  We have emphasized this latter point by demonstrating that chain-flips are {\it necessary} for the MCMC simulation to find the proper ground state in the case where one of the members of the extensive manifold is made to have lower energy.  In this perturbed model, chain-flips promote a low temperature phase transition to a long-range ordered state, that recovers all of the residual entropy of the unperturbed model.

We have also used our MCMC algorithm to study the physically important extension of the FF honeycomb Ising model, where an external magnetic field $h$ is applied.  In this case, moderate values of $h$ promote the realization of partially-ordered states, corresponding to magnetization plateaus with values of $M=1/8$ and $M=1/4$.  The precise nature of the partially-ordered states is dependent on the geometry with which frustration is introduced into the original (unperturbed) FF honeycomb Ising model, and is not gauge-invariant in this sense.  An interesting phenomenon that occurs is the reemergence of extensive entropy ``spikes'' at $h$ values bounding the $M=1/4$ plateau.  Using a proof based on the reduction of the configurational disorder down to a Fibonacci recurrence, we are able to show that the entropy of the two reemergent spikes is equal to 
\begin{eqnarray}
S/N&=& \ln \phi / 8\approx0.06015 \hspace{.5cm} {\rm for} \hspace{3mm} h/J=1, \nonumber \\
S/N&=& \ln \phi / 2\approx0.24061 \hspace{.5cm} {\rm for} \hspace{3mm} h/J=3/2. \nonumber
\end{eqnarray}
where $\phi \approx 1.618$ is the golden ratio.  
MCMC simulations confirm these results to a high degree of accuracy.
It is interesting to note that, in one case
 (for $h/J = 3/2$), the entropy spike $S/N = 0.241$ is actually greater than the residual entropy for $h=0$ ($S/N=0.214$).
The phenomenon of a magnetization plateau bounded by extensive entropy spikes has previously been seen to occur on several models in one \cite{1dS} and two dimensions \cite{triang_h}, and also in 3D spin ice systems in an applied field along the [111] crystallographic direction \cite{Isakov}.

\section{Acknowledgements}

We are indebted to P. Holdsworth, M. Hastings and A. Burkov for enlightening discussions.  
R.G.M. is grateful for an ATOSIM fellowship from the \'Ecole Normale Sup\'erieure de Lyon, where some of these ideas were developed.
Support for this work was provided by NSERC of Canada.

\appendix
\section{Number of $m$-digit binary numbers with $d$ transitions from 0 to 1.}
Let $R(m,d)$ be the number of (periodic) $m$-digit binary numbers with $d$ transitions from 0 to 1. We want to show that
\begin{equation}
R(m,d)=2\left(\begin{array}{c} m\\2\, d\end{array}\right).
\label{Rmd}
\end{equation}
In a periodic binary number of size $m$, there are $m$ possible places to switch digits.
Choose $2d$ of these $m$ places in order to get $d$ transitions from 0 to 1.
Once the $2d$ locations where digits switch are chosen, the number can be formed in two different ways (zeros and ones can be switched). This leads directly to expression (\ref{Rmd}).
\section{Number of configurations $g(n)$.}

Consider a (periodic) line with $n$ spins. We want to show that $g(n)$, the number of configurations in which no two adjacent spins are down, satisfies the Fibonacci recurrence equation,
\begin{equation}
g(n)=g(n-1)+g(n-2),
\label{eq:fibo}
\end{equation}
with exact solution
\begin{equation}
g(n)=c_1\,\phi^n+c_2\,\theta^n,
\label{eq:FibSol2}
\end{equation}
where
\begin{align}
\phi&={1+\sqrt{5} \over 2}\approx1.6180, \nonumber\\
\theta&={1-\sqrt{5} \over 2}\approx-0.6180,\nonumber\\
c_1&={5+3\,\sqrt{5}\over 10}\approx1.1708,\label{eq:constants}\\
c_2&={5-3\,\sqrt{5}\over 10}\approx-0.1708.\nonumber
\end{align}

Note that this also covers the case of a (periodic) line with $n$ fixed spin pairs, in which no two adjacent spin pairs are down.

To count the number of valid states, we introduce two sets.  First, $U_n$ is the set of valid states on rows of size $n$ such that the first spin in the row is an up spin.  We would like to find $|U_n|$ recursively.  We divide $U_n$ into two categories, rows starting with two up spins and rows starting with an up spin and then a down spin.  An up spin followed by any member of $U_{n-1}$ is in the first category, and all members of the first category must be of that form.  If the row starts with an up spin and then a down spin, then the third spin must be an up spin, so the second category is every row of the form up spin, down spin, then a member of $U_{n-2}$.  So
\begin{align}
  U_n = (\uparrow + U_{n-1}&) \quad \cup \quad (\uparrow + \downarrow + U_{n-2}) \nonumber\\
  \nonumber\\
  \Rightarrow \quad |U_n| &= |U_{n-1}| + |U_{n-2}|.
  \label{Up_Set}
\end{align}

Now our second set is $T_n$, the set of all valid states on rows of size $n$.  Clearly $U_n \subset T_n$, so we need only worry about rows that begin with a down spin.  If a row begins with a down spin, its second spin must be an up spin.  Furthermore, since the row loops around due to the periodicity of the lattice, its last spin must be an up spin.  Thus rows that begin in a down spin, end in an up spin, and have a member of $U_{n-2}$ in between to account for all remaining members of $T_n$.  So
\begin{align}
  &T_n = U_n \quad \cup \quad (\downarrow + U_{n-2} + \uparrow) \nonumber \\ 
  \nonumber\\
  \Rightarrow \quad|T_n| &= |U_n| + |U_{n-2}| \nonumber \\
   &= |U_{n-1}| + |U_{n-2}| + |U_{n-3}| + |U_{n-4}| \text{, by \eqref{Up_Set}} \nonumber \\
   &= |T_{n-1}| + |T_{n-2}| \label{T_Set} \nonumber
\end{align}

With $|T_n|=g(n)$, we obtain the Fibonacci equation, Eq.\ (\ref{eq:fibo}), for $g(n)$.
The general solution of this recurrence is given by Eq.\ (\ref{eq:FibSol2}), with $\phi$ and $\theta$
the roots of the characteristic polynomial of the equation.
The base cases (by inspection) are $g(1) = 2$ and $g(2) = 3$, which provide the values for the constants $c_1$ and $c_2$ that are given in (\ref{eq:constants}).

\section{Closed-form expression for Eq.\ (\ref{eq:sum}).}
We show that
\begin{equation}
\Omega=\sum_{d=0}^{m/2} R(m,d)\,\alpha^d,
\end{equation}
with $R(m,d)$ given by Eq.\ (\ref{eq:Rmd}),
has the closed-form expression
\begin{equation}
\Omega=\left(1+\sqrt{\alpha}\right)^{m}+\left(1-\sqrt{\alpha}\right)^{m}.
\label{eq:omega3}
\end{equation}
We show this assuming that $m$ is even ($m=L/2$ in Eq.\ (\ref{eq:sum}), and $L$ is a multiple of 4.)
Using
\begin{align*}
(1+\sqrt{\alpha})^m&=\sum_{i=0}^{m}\left(\begin{array}{c} m\\i\end{array}\right)(\sqrt{\alpha})^i\\
&=\sum_{i=0}^{m/2}\left(\begin{array}{c} m\\2i\end{array}\right)(\sqrt{\alpha})^{2i}\\
&+\sum_{i=0}^{m/2-1}\left(\begin{array}{c} m\\2i+1\end{array}\right)(\sqrt{\alpha})^{2i+1},
\end{align*}
and
\begin{align*}
(1-\sqrt{\alpha})^m
&=\sum_{i=0}^{m}\left(\begin{array}{c} m\\i\end{array}\right)(-\sqrt{\alpha})^i\\
&=\sum_{i=0}^{m/2}\left(\begin{array}{c} m\\2i\end{array}\right)(\sqrt{\alpha})^{2i}\\
&-\sum_{i=0}^{m/2-1}\left(\begin{array}{c} m\\2i+1\end{array}\right)(\sqrt{\alpha})^{2i+1},
\end{align*}
we obtain immediately
\begin{equation}
\Omega=2\, \sum_{d=0}^{m/2} \left(\begin{array}{c} m\\2\, d\end{array}\right)\,\alpha^d=
\left(1+\sqrt{\alpha}\right)^{m}+\left(1-\sqrt{\alpha}\right)^{m}.
\end{equation}

\bibliography{FF_biblio}

\end{document}